\documentclass[useAMS,plainnat]{mn2e}

\usepackage{psfig,graphicx,epsfig}

    \newcommand{\Msun}{$M_{\odot}$}

\newcommand{\snr}{{W49B}}

\title[overionized plasma in W49B]{Unveiling the spatial structure of the overionized plasma in the supernova remnant W49B}
\author[Zhou et al.]{Xin Zhou$^{1,2}$,
 Marco Miceli$^{3,2}$, Fabrizio Bocchino$^{2}$, Salvatore Orlando$^{2}$, and Yang Chen$^{1,4}$\\
$^{1}$Department of Astronomy, Nanjing University, Nanjing 210093, China\\
$^{2}$INAF-Osservatorio Astronomico di Palermo, Piazza del Parlamento 1, 90134 Palermo, Italy\\
$^{3}$Dip. di Scienze Fisiche \& Astronomiche, Univ.\ di Palermo, Piazza del Parlamento 1, 90134 Palermo, Italy\\
$^{4}$Key Laboratory of Modern Astronomy and Astrophysics (Nan-jing University), Ministry of Education, China
}
\begin{document}

\date{Accepted. Received}

\pagerange{\pageref{firstpage}--\pageref{lastpage}} \pubyear{}

\maketitle

\label{firstpage}

\begin{abstract}

W49B is a mixed-morphology supernova remnant with thermal X-ray emission
dominated by the ejecta. In this remnant, the
presence of overionized plasma has been directly established, with
information about its spatial structure. However, the physical origin
of the overionized plasma in W49B has not yet been understood. We
investigate this intriguing issue through a 2D hydrodynamic model
that takes into account, for the first time, the mixing of ejecta with
the inhomogeneous circumstellar and interstellar medium, the thermal
conduction, the radiative losses from optically thin plasma, and the
deviations from equilibrium of ionization induced by plasma dynamics. The
model was set up on the basis of the observational results. We found that
the thermal conduction plays an important role in the evolution of W49B,
inducing the evaporation of the circumstellar ring-like cloud (whose
presence has been deduced from previous observations) that mingles
with the surrounding hot medium, cooling down the shocked plasma,
and pushes the ejecta backwards to the center of the remnant, forming
there a jet-like structure. During the evolution, a large region of
overionized plasma forms within the remnant. The overionized plasma
originates from the rapid cooling of the hot plasma originally heated
by the shock reflected from the dense ring-like cloud. In particular,
we found two different ways for the rapid cooling of plasma to appear:
i) the mixing of relatively cold and dense material evaporated from the
ring with the hot shocked plasma and ii) the rapid adiabatic expansion of
the ejecta. The spatial distribution of the radiative recombination continuum
predicted by the numerical model is in good agreement with that observed.
\end{abstract}

\begin{keywords}
hydrodynamics -- ISM: individual object: W49B, G43.3$-$0.2 -- ISM:
supernova remnants -- methods: numerical.
\end{keywords}

\section{Introduction}

W49B is one of the brightest Galactic supernova remnant (SNRs)
in the radio (Moffett \& Reynolds 1994) and X-ray bands (Immler \&
Kuntz 2005). It exhibits incomplete radio shell with centrally filled
thermal X-rays (Pye et al.\ 1984; Smith et al.\ 1985), which is the
characteristic of mixed-morphology supernova remnants (MM SNRs) (Rho \&
Petre 1998). However, at odds with typical MM SNRs, its X-ray emission
is ejecta-dominated (Smith et al.\ 1985; Fujimoto et al.\ 1995; Hwang
et al.\ 2000) and presents a central jet-like feature and a bright limb
to the East (Miceli et al.\ 2006; Keohane et al.\ 2007). Near-infrared
narrow-band observations revealed a barrel-shaped structure with coaxial
rings (Keohane et al.\ 2007), probably a remnant of a bipolar wind
surrounding the massive progenitor star.
The X-ray jet-like structure
is along the axis of the barrel. A strip of shocked molecular hydrogen
was also found in the eastern region just outside the shell, indicating
that the remnant is interacting with a molecular cloud there.

Recently, a strong radiative recombination continuum (RRC) associated
with H-like Fe was detected in the {\sl SUZAKU} X-ray spectrum of W49B
by Ozawa et al.\ (2009), thus showing the direct
signature of overionization in the ejecta. Miceli et al.\ (2010)
confirmed the RRC features by a spatially resolved spectral analysis
of the {\sl XMM-Newton} data, and found that the overionized plasma
is localized in the center of the remnant and in the western region,
but not in the eastern region.

The origin of such overionized recombining plasma is not well understood
and rather counter-intuitive in objects like SNRs where we mostly
expect underionized plasma. Kawasaki et al.\ (2002) proposed that the
hot interior of a SNR can, under some circumstances, cool rapidly via
thermal conduction with the cooler exterior, but Yamaguchi et al.\ (2009)
pointed out some difficulties for this scenario. 
By analysing the observations of IC 443, Yamaguchi et al. (2009)
reexamined the time scale of thermal conduction that account for
overionization in Kawasaki et al. (2002), and found that it is far longer
than the remnant age.  The thermal conduction alone therefore cannot
account for the observed overionized plasma in IC 443, whose origin
therefore remain still not well understood.

However, the estimate of Yamaguchi et al. (2009) is based on the
temperature difference between hot and cold X-ray emitting component. It
may still be possible (and indeed very probable) that surrounding
ISM/CSM inhomogeneities (in the form of rings, large molecular clouds,
cavity walls, etc.), once shocked, will give rise to a conductive
thermal flux between the shocked molecular material and the inner part
of the SNR. This may be indeed the case of W49B, where the higher inner
temperatures than IC 443 (due to the lower age of the remnant) may
further increase the effects of thermal conduction with shocked
inhomogeneities.

To explain the morphology and chemical structure of W49B, two different
scenarios have been proposed: either a bipolar explosion of the former
supernova or a spherical supernova explosion expanding through a strongly
inhomogeneous surrounding medium (Miceli et al.\ 2006, 2008, Keohane et
al.\ 2007, Lopez et al.\ 2009). Most of these authors leave both scenarios
open, but Lopez et al.\ (2009) strongly supports the bipolar explosion
scenario, on the base of the Fe abundances measured with CIE spectral
analysis in several regions of W49B. However, as shown by Ozawa et al.\
(2009), the abundances are different if using either CIE or overionization
model. As a result, the conclusions of Lopez et al.\ (2009) may depend
on the CIE model they used. In the light of these considerations, we
believe that there is no observational evidence, at the present time, to
prefer one of the two scenarios discussed above; the case of a spherical
explosion expanding through an inhomogeneous medium is still realistic
if considering that the plasma may be in non-equilibrium of ionization.
In this paper, we focus on the reproduction of the morphology and
overionization pattern of W49B by adopting the simpliest SNR model,
describing a spherical explosion. We will explore the alternative
scenario, namely the bipolar explosion, in a forthcoming paper.

\begin{figure*}
\centerline{ {\hfil\hfil \psfig{figure=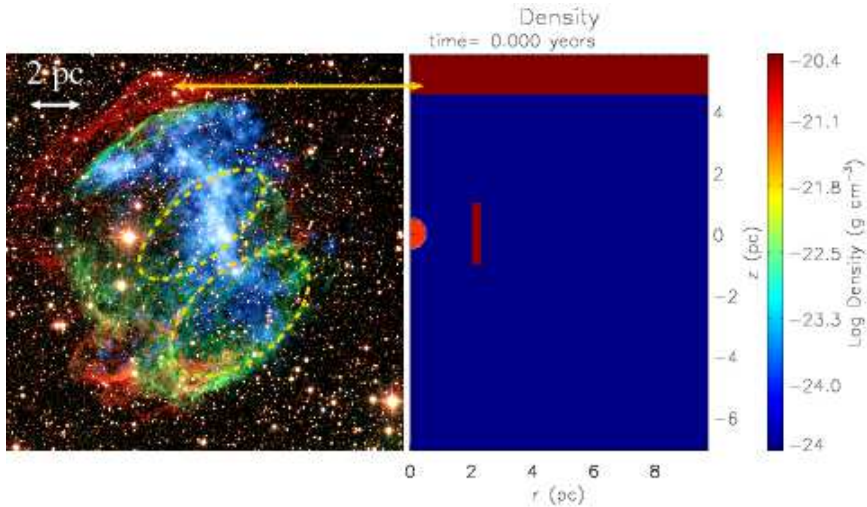,height=7.0cm,angle=0,clip=} \hfil\hfil}}
\caption{($left$): Color composite image of W49B: 2.12 $\mu$m $H_2$
(red), 1.64 $\mu$m [Fe {\sc ii}] (green), and $Chandra$ X-ray (blue)
(see also Miceli et al.\ 2006 and Keohane et al.\ 2007). The image has been rotated so that East is up and North is to the right.  ($right$):
Initial conditions of the simulation. The section in the $(r,z)$ plane of
the density distribution is shown. The model consists of a spherical SNR,
a dense ring (corresponding to the coaxial rings shown in green
in the left panel, and indicated by yellow ellipses), and an
upper dense cloud (corresponding to the eastern molecular cloud shown in red
in the left panel; the correspondence is indicated by the yellow
arrows). The symmetry axis in cylindrical coordinates is at $r=0$.
}
\label{f:init}
\end{figure*}

In this paper, we introduce a hydrodynamic model whose initial conditions
were set up to describe appropriately W49B and investigate 
the physical
origin of the peculiar morphology of the remnant and of the regions of
overionization recently detected. The numerical model and the results are
described in Sect.~2 and Sect.~3, respectively; the conclusions are
summarized in Sect.~4.

\section{Hydrodynamic modeling}

Our model describes the evolution of an originally spherical SNR expanding
through an inhomogeneous medium and, in particular, interacting with
the surrounding dense clouds. The remnant evolution is modeled by
numerically solving the time-dependent equations of mass, momentum,
and energy conservation (as described in Orlando et al.\ 2005),
taking into account the effects of thermal conduction (including the
effects of heat flux saturation) and radiative losses from an optically
thin plasma in collisional ionization equilibrium (CIE). In addition,
the model includes the deviations from equilibrium of ionization induced
by the dynamics as described in Reale \& Orlando (2008). Although our
description is not entirely self-consistent (since radiative losses are
obtained assuming CIE), this does not significantly affect our results,
because energy losses are dominated by transport by thermal conduction
at all critical evolutionary phases (e.g.\ the interaction of the shock
with the dense clouds, see Sect.~3) and the radiative time-scales are
much larger than the age of the remnant.

We performed the calculations in a 2-D cylindrical coordinate system
$(r,z)$ using the FLASH code v2.5 (Fryxell et al.\ 2000), and including
additional computational modules to handle the thermal conduction, the
radiative losses, and the non-equilibrium of ionization (NEI) effects 
(see Orlando et al.\ 2005 and Reale
\& Orlando 2008 for the details of their implementation). We assumed
that the fluid is fully ionized, and can be regarded as a perfect gas
(with a ratio of specific heats $\gamma = 5/3$). We traced the ejecta
material by considering a passive tracer initially set to 1 for pure
ejecta, and 0 elsewhere.

The right panel of Fig.~\ref{f:init} shows the initial configuration
of our model in the $(r,z)$ plane. The setup was dictated by the results
of the analysis of the X-ray and infrared observations of W49B (shown,
for comparison, in the left panel of Fig.~\ref{f:init}). We started
our simulation considering a spherical remnant with radius\footnote{
The initial radius of the remnant has been chosen
in order to have an optimal trade-off between computational time,
code stability, and accuracy. In particular, we have verified that the
expansion relaxes to the self-consistent free-expansion phase before
the remnant hits the cloud ring, thus ensuring that our choice of 0.5
pc is optimal.} 0.5 pc. In
agreement with Miceli et al.\ (2008), the total mass of ejecta was set
to 6 {\Msun} and the explosion energy to $10^{51}$ erg (partitioned so that
$\sim99.99\%$ of the SN energy is kinetic energy). 
It is worth noting that the mass of ejecta derived from the observations
is subject to some uncertainties, given, for instance, the uncertainty
in converting emission measure to mass. In our model, a different mass
of the ejecta implies a different velocity of the ejecta (if
keeping the explosion energy constant), thus influencing the energy of the
shocks reflected by the clouds. This point may be important because,
as shown below, the reflected shocks determine both the morphology and
overionization pattern of the remnant. However, the ejecta
velocity is $v_{\rm ej} \propto m_{\rm ej}^{-1/2}$ (where $m_{\rm ej}$
is the ejecta mass), so that we do not expect dramatic changes in our
results if considering the uncertainty on the ejecta mass to be within
a factor of 2.
The ambient medium in which the
remnant expands is assumed to be inhomogeneous. In particular, we focused
on the coaxial rings observed by Keohane et al.\ (2007) in 1.64 $\mu$m [Fe
{\sc ii}], suggesting that \snr\ is confined by ring-like enhanced density structures.
These structures may originate from the bipolar wind
of the massive progenitor star as indicated by Keohane et al.\ (2007), still, there
are another two possibilities: the protostellar
disk that had been evaporated by the ionizing radiation of the progenitor
(McCray \& Lin 1994), or the accumulated material in the orbiting plane in a binary system.
These coaxial rings are also visible in the radio continuum
band (Lacey et al.\ 2001) and may also be the origin of the X-ray emission
in the remnant (Miceli et al.\ 2006; Keohane et al.\ 2007). According
to this barrel-shaped structure, a dense ring surrounding the remnant
(2 pc away from the central axis) was set up in our model (its section
in the $(r,z)$ plane is shown in the right panel of Fig.~\ref{f:init}
as a vertical bar). We fixed its density to 1600 cm$^{-3}$ in agreement
with the results of Keohane et al.\ (2007). In addition, we accounted for
the molecular H$_2$ cloud observed in the eastern region of the remnant,
by modeling a dense wall (density $n=2000$ cm$^{-3}$, Keohane et al.\
2007; see horizontal bar in the right panel of Fig.~\ref{f:init}). 
It is worth emphasizing that the density values used in our
model are those constrained by observations and may be affected by some
uncertainties. Nevertheless we do not expect any change in our results
as long as these density values remain few orders of magnitude higher
then the intercloud medium. In such a case, the clouds act as walls,
leading to strong reflected shocks.
The tenuous intercloud medium was assumed to be isothermal ($T_{\rm ism} =
10^4$ K) and homogeneous (density $n_{\rm ism} = 0.1~{\rm cm}^{-3}$).
Both the dense ring and the molecular cloud are assumed to be in pressure
equilibrium with the  intercloud medium. 
Note that our model is aimed at reproducing the centrally-peaked
morphology of the remnant and its pattern of overionization.

The computational domain is $\sim 10\times 13$ pc. At the coarsest
resolution, the adaptive mesh algorithm (PARAMESH; MacNeice et al.\ 2000)
included in the FLASH code uniformly covers the 2D computational domain
with a mesh of $3\times4$ blocks, each with $8^2$ cells. At the beginning
of the simulation we allowed for 7 additional nested levels of refinement
with resolution increasing twice at each refinement level. During the
remnant evolution, the number of nested levels progressively decreased
down to 5 at the end of the simulation. This grid configuration yields
an effective resolution at the finest level of $\sim 0.003$ pc at the
beginning and 0.012 pc at the end of the simulation.  Reflecting boundary
conditions were used at $r = 0$ pc, consistently with the adopted
symmetry; free outflow conditions were used elsewhere.

\section{Results}

\subsection{Hydrodynamical evolution}

\begin{figure}
\centerline{ {\hfil\hfil \psfig{figure=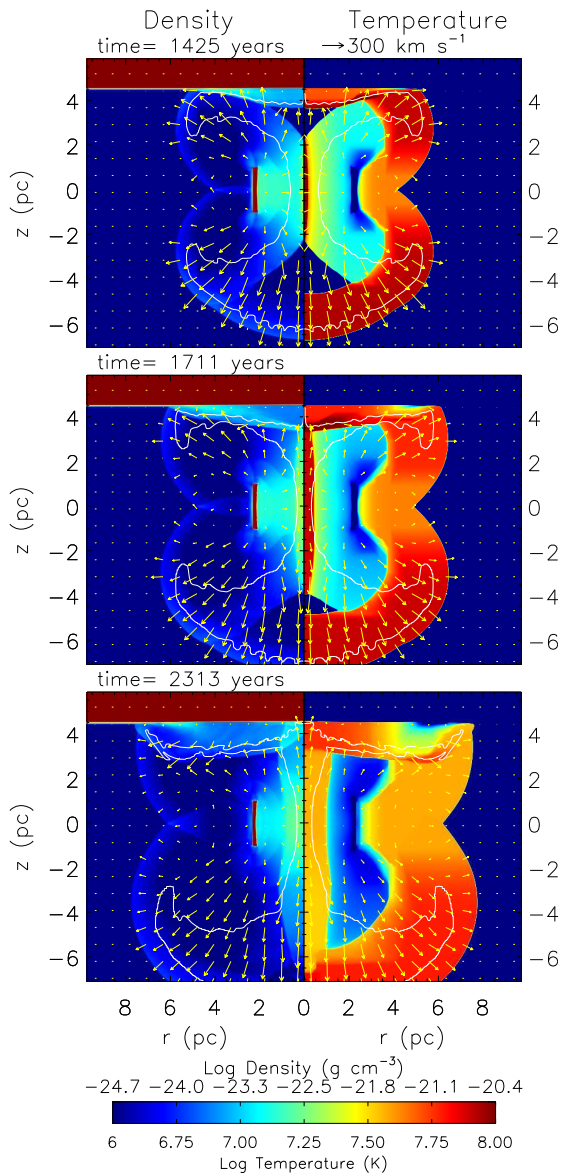,height=15.5cm,angle=0,clip=} \hfil\hfil}}
\caption{2-D sections in the $(r,z)$ plane of the mass density
   distribution (left) and temperature (right), both in log scale,
   at three different evolutionary stages (see upper left corner of
   each panel). The velocity arrows scale linearly with respect to the
   reference velocity shown in the upper right corner of the upper panel
   and corresponding to 300 km s$^{-1}$.}
\label{f:dentemp}
\end{figure}

Fig.~\ref{f:dentemp} shows the spatial distributions of density and
temperature in the $(r,z)$ plane at three different evolutionary stages;
the white contours enclose the ejecta material (i.e. the fluid cells
whose content is ejecta by more than $90\%$). During the first phase
of evolution, the remnant expands through the homogeneous intercloud
medium impacting on the ring-like cloud at $t\sim 250$ yr and on the
molecular cloud at $t\sim 790$ yr (not shown in Fig.~\ref{f:dentemp}). After the interaction of the forward
shock with the dense ring and the eastern dense cloud, reflected shocks
develop and move backwards to the center of the remnant. At the same
time, the cold cloud material evaporates under the effect of the thermal
conduction, mixing with the surrounding hot shocked plasma.  In the time
lapse between $t\sim 1400$ yr and $t\sim 1700$ yr (upper and middle panel
of Fig.~\ref{f:dentemp}), the hot plasma around the circumstellar ring
rapidly cools down because of the mixing of the hot shocked plasma with
the dense and cooler plasma evaporated from the ring. After $t\sim 1700$
yr (middle and lower panel of Fig.~\ref{f:dentemp}), the ejecta heated by
the reflected shock also rapidly cool down due to their rapid adiabatic
expansion\footnote{In the central region of the remnant this adiabatic
cooling is masked by the re-heating of ejecta by reflected shocks
``focusing" to the center.}. The lower panel in Fig.~\ref{f:dentemp} also
shows that at $t\sim 2300$ yr the expansion of the ejecta in the east 
direction is hampered by the presence of the dense molecular cloud. On the
contrary, the ejecta expand freely westwards, progressively cooling down.

Using Eqs.
(10) and (11) in Orlando et al. (2005), we estimated the time-scales for
thermal conduction and radiative losses to be a few $10^3$ yr and a few
$10^6$ yr, respectively, in most of the intercloud region of the remnant
at $t = 1400$ yr. At later evolutionary stages, the results are the
same. We conclude, therefore, that the energy losses are dominated by
transport by thermal conduction at all critical evolutionary phases,
and the radiative scales are much larger than the age of the remnant.
The low conduction timescale are due to the large temperature contrast
between shocked CSM/MC material and the inner part of the remnant,
which was not considered in the IC 443 by Yamaguchi et al. (2009).

The reflected shock from the dense ring concentrates to the center of
the remnant, squeezing the ejecta there, and forming a jet-like feature
with dense and hot material. We verified (with the help of an additional
simulation with the same setup of that presented here but neglecting the effects of thermal conduction) that, if the thermal conduction is not included in the
simulation, the ejecta are distributed over a widespread region and are
not pushed backwards to the center of the remnant; as a consequence if it were not for the thermal conduction, the
central jet-like feature would be too tenuous to be distinguished. At
$t\sim 2300$ yr, the central jet-like feature reaches the eastern dense
cloud, and a dense and hot eastern shell (rich of ejecta) forms. The central
``jet'' region and the eastern ``shell'' region are hot enough to emit X-rays,
and their surface brightness is expected to be high because of their high
density and enhanced abundances. At this stage of evolution, therefore,
our hydrodynamic model naturally reproduces the central jet-like feature
and the eastern bright shell observed in X-rays with \emph{XMM-Newton}
and $Chandra$ (Miceli et al.\ 2006; Keohane et al.\ 2007; Lopez et
al.\ 2009). The thermal conduction and the surrounding clouds are key
ingredients to reproduce the observed morphology. Since at $t\sim 2300$
yr both the morphology and the level of overionization (see Section 3.3)
are in good agreement with X-ray observations, we suggest that this
evolutionary stage can provide a good estimate for the age of W49B.

\subsection{Deviations from equilibrium ionization}

\begin{figure}
\centerline{ {\hfil\hfil \psfig{figure=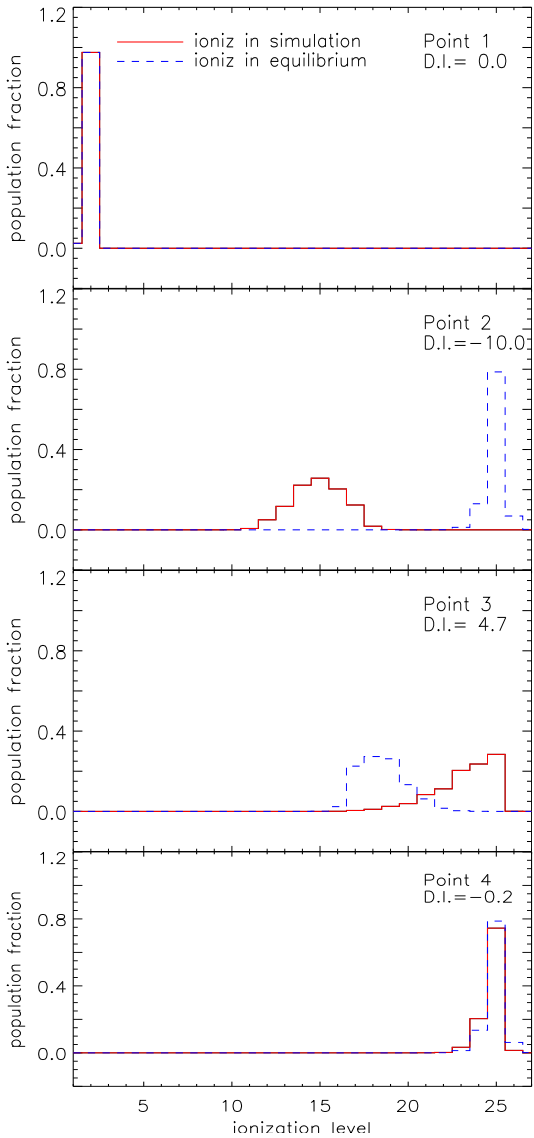,height=15cm,angle=0,clip=} \hfil\hfil}}
\caption{Ionization population plots of iron extracted from the
four points indicated in the bottom panel of Fig.~\ref{f:nei}.}
\label{f:pop} \end{figure}

\begin{figure}
\centerline{ {\hfil\hfil \psfig{figure=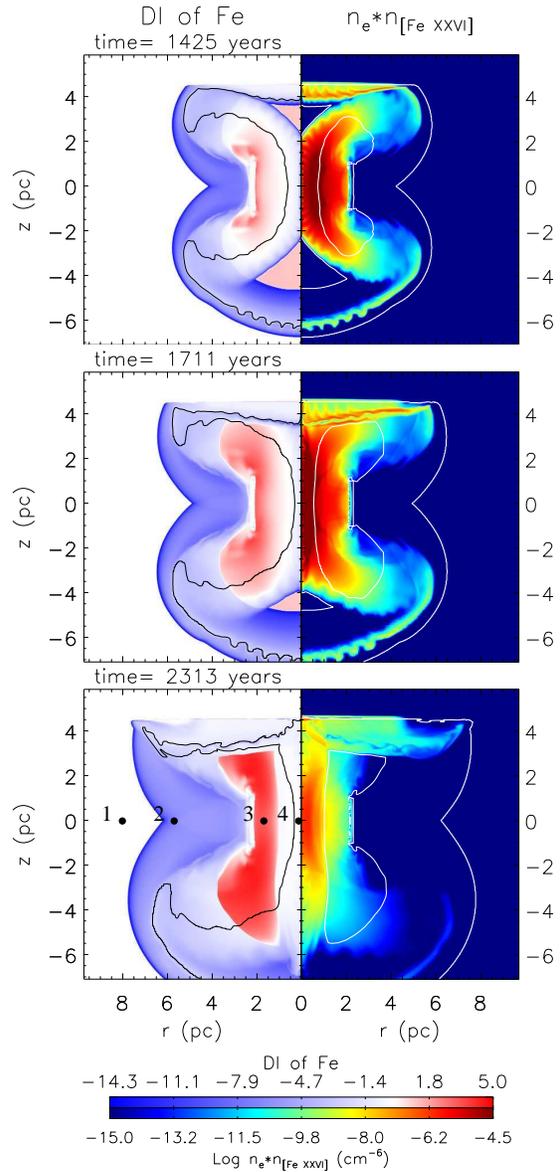,height=15.5cm,angle=0,clip=} \hfil\hfil}}
\caption{2-D sections in the $(r,z)$ plane of the differential
ionization of iron (left; see the text for the definition of DI) and
the emission measure of [Fe {\sc xxvi}] RRC (right), at the
three evolutionary stages as in Fig.~\ref{f:dentemp} (see upper left
corner in each panel). The black contours (on the left) enclose the
ejecta material, the white contours (on the right) mark the value DI =
0. Positive values of DI (marked in red in the left panels) indicate
the case of overionization; negative values (blue) indicate the case
of underionization. In the calculation of the emission measure of
[Fe {\sc xxvi}] line emission, we have assumed that the metal abundance of iron is 5 Z$_{\sun}$, according to Miceli et al.\ (2006).
The circular points in the third panel are the points selected for showing
the ionization population in Fig.~\ref{f:pop}.}
\label{f:nei}
\end{figure}

In this section, we address the physical origin of the overionized H-like
Fe detected in W49B by Ozawa et al.\ (2009) and localized in the center
of the remnant and in the western region, but not in the eastern region
(Miceli et al.\ 2010). We examine the distribution of
underionized/overionized plasma, by defining the iron Differential
Ionization (DI) as

\begin{equation}
{\rm DI} =
\left<l\right>_{\rm simulation} - \left<l\right>_{\rm
equilibration}~,
\label{eq:DI}
\end{equation}

\noindent
\[
{\rm where}~~~~
\left<l\right> = \frac{\Sigma (f_i\times l_i)}{\Sigma (f_i)}
\]

\noindent
is the average ionization level weighted by the population fraction,
$f_i$ is the population fraction of the $i$th ion of Fe, $l_i$
is the ionization level of the $i$th ion, and the sum is performed
over all the Fe ionization stages. The subscripts ``simulation" and
``equilibration" in Eq.~\ref{eq:DI} refer to the value obtained in
a given cell of the simulation domain and to that expected in CIE,
respectively. Fig.~\ref{f:pop} shows the population fractions of Fe
versus the ionization levels derived in four different computational
cells of our domain (in red, see Fig.~\ref{f:nei} for the location of the cells), compared to those expected in CIE, at the same location
(in blue). The DI parameter is a reliable indicator to distinguish the case of
underionization/overionization: positive values indicate overionization
(as in point \#3 in Fig.~\ref{f:pop}), negative values indicate
underionization (as in point \#2 in Fig.~\ref{f:pop}).

The left panels of Fig.~\ref{f:nei} show the spatial distribution of
the Fe DI at the three evolutionary stages of Fig.~\ref{f:dentemp},
with black contours marking the ejecta overlaid. The right panels show
the corresponding distributions of emission measure of the [Fe {\sc xxvi}]
RRC (the white contours mark the value DI = 0). At $t\sim
1400$ yr, the overionized material (positive DI values, marked in red)
is around the dense ring, and originates from the plasma previously heated by the
reflected shock and rapidly cooling down because of the mixing with the
cooler material evaporating from the dense ring (see Section 3.1 for a
detailed description of the evolution of the system). At $t\sim 1700$ yr
(middle panel of Fig.~\ref{f:nei}), conditions of overionization arise
far from the ring. In this case the overionization is associated with the
adiabatic cooling of the rapidly expanding ejecta (both in the east and
in the west directions). At $t\sim 2300$ yr the adiabatic expansion of the
ejecta in the east direction is hampered by the dense molecular cloud;
the plasma there does not cool down anymore but is re-heated due to
the interaction with the shock reflected by the molecular cloud. As
a consequence, the conditions of overionization there are not present
anymore. On the contrary, they are still present in the western region
where the plasma keeps to expand freely. Since the time $t\sim
1700$ yr and onwards, the overionized plasma is composed by both shocked
ISM and ejecta, as it is shown by the intersection of the red region
and the ejecta contours in the left middle and left bottom panels of
Fig.~\ref{f:nei}.

\subsection{Strength of radiative recombination continuum (RRC)}

We compare our model results with X-ray observations of W49B by computing,
for each computational cell, the ratio $\rho$ between the X-ray luminosity
of the [Fe {\sc xxvi}] RRC and the thermal bremsstrahlung emission. The
diagnostic power of this ratio consists in getting the information about the
strength of the RRC. In fact, high values of $\rho$ qualitatively
corresponds to regions where the RRC edge has the highest probability
of being observed. According to the radiative recombination model in
ATOMDB (see Smith \& Brickhouse 2002) v1.3.1, the luminosity of the [Fe {\sc xxvi}] RRC is

\begin{equation}
L_{\rm RRC} \propto n_e n_{[\rm Fe~{\sc XXVI}]}
\exp\left(-\frac{E_\gamma-E_{\rm edge}}{k T}\right)T^{-3/2}
\end{equation}

\noindent
where $E_{\rm edge}$ is the K-edge energy of the [Fe {\sc xxv}] which is
8.83 keV and $E_\gamma$ is the energy
of the emitted photon which is set to be equal to $E_{\rm edge}$ in our strength examination.
The approximation of the luminosity of thermal bremsstrahlung emission is
proportional to (Longair 1994):

\begin{equation}
\left\{
\begin{array}{c}
\displaystyle
T^{-1/2} n_{\rm e} n_{\rm i} log(\frac{k T}{E_\gamma}) exp(-\frac{E_\gamma}{k T})$,
~~~~when $E_\gamma < k T \\
\\
\displaystyle
T^{-1/2} n_{\rm e} n_{\rm i} (\frac{E_\gamma}{k T})^{1/2} exp(-\frac{E_\gamma}{k
T})$, ~~~~when $E_\gamma \gg kT ~.\\
\end{array}
\right.
\end{equation}

\noindent
The ratio $\rho$ between the luminosity of the [Fe {\sc xxvi}] RRC and
the thermal bremsstrahlung emission is proportional to:

\begin{equation}
\left\{
\begin{array}{c}
\displaystyle
\frac{n_{\rm [Fe~{\sc XXVI}]}}{n_{\rm i}} T^{-1} log(\frac{k
T}{E_\gamma})^{-1} exp(\frac{E_{\rm edge}}{k T})$, when $E_\gamma < k T \\
\\
\displaystyle
\frac{n_{\rm [Fe~{\sc XXVI}]}}{n_{\rm i}} T^{-1} (\frac{k
T}{E_\gamma})^{1/2} exp(\frac{E_{\rm edge}}{k T})$, when $E_\gamma
\gg k T~. \\
\end{array}
\right.
\end{equation}

\begin{figure}
\centerline{ {\hfil\hfil \psfig{figure=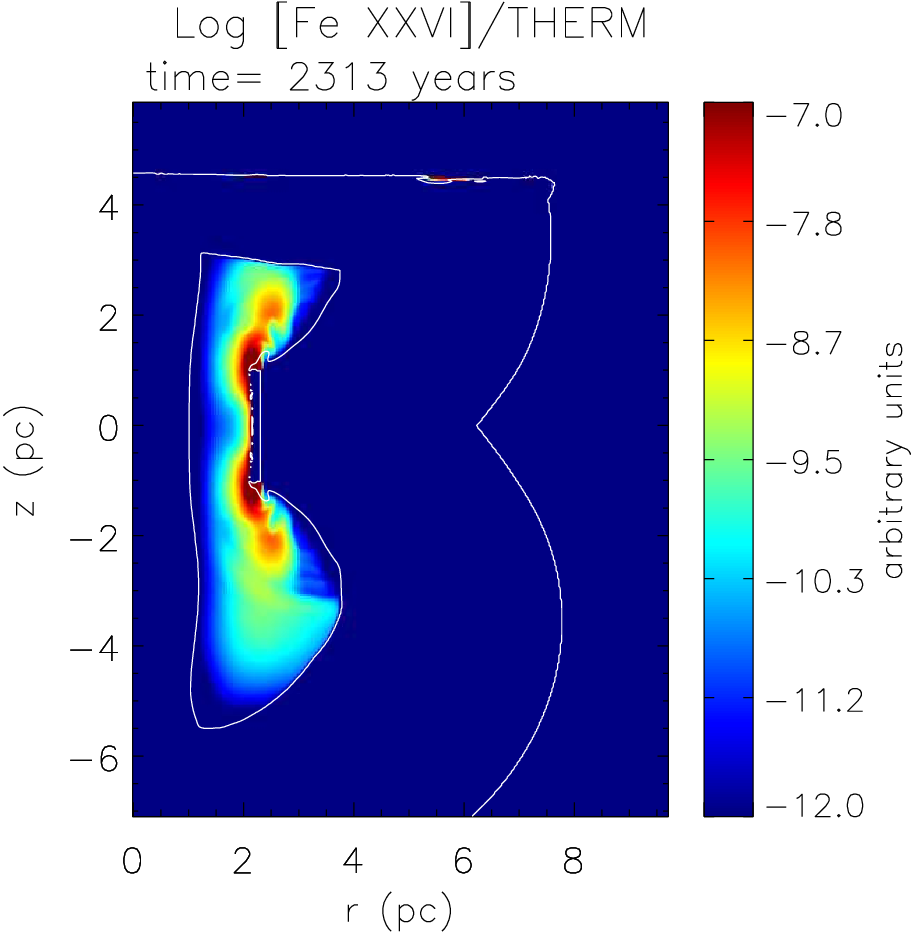,height=7.5cm,angle=0,clip=} \hfil\hfil}}
\caption{Map of the ratio between the luminosities of RRC of [Fe {\sc xxvi}]
and thermal bremsstrahlung emission in the $(r,z)$ plane, with the contour marking the value DI = 0 the same
as the contour in the right half of each panel in Fig.~\ref{f:nei}. The luminosities of RRC of [Fe {\sc xxvi}] and thermal bremsstrahlung emission are in arbitrary units. The
ejecta material is considered to be with the abundance of 5 Z$_{\sun}$, corresponding to that found by Miceli et al.\ (2006). The larger the ratio the more visible the RRC.}
\label{f:RRC}
\end{figure}

To compute $\rho$ we also assume a pure ejecta plasma which can be
mimicked by setting the abundances to high values, we explored the Fe
abundance in the range of 1--10000 Z$_{\sun}$, while we assumed solar
abundance for ISM and CSM. We adopted the ejecta tracer as a factor to
evaluate the real Fe abundance in the computational cells where ejecta
and ISM are mixed together. We found that different choices of the
abundance of pure ejecta have only secondary effects on $\rho$. In
particular, the [Fe {\sc xxvi}] RRC is already clearly visible at 5
Z$_{\sun}$ (the value found by Miceli et al.\ (2006) for W49B),
as it is shown by the map of
$\rho$ in the (r,z) plane at $t\sim2300$ yr presented in Fig.~\ref{f:RRC}, and it
is more visible with larger abundance.
The [Fe {\sc xxvi}] RRC is mainly visible at the center and in the
western region (corresponding to the western part of W49B), in good
agreement with the observations (Miceli et al.\ 2010).

\section{Summary and conclusions}

We modeled the evolution of W49B taking into account, for the first
time, the mixing of ejecta with the inhomogeneous circumstellar and
interstellar medium, the thermal conduction, the radiative losses from
optically thin plasma, and the deviations from equilibrium of ionization
induced by the dynamics. Our model is suited to address the still pending
issue of the origin of overionized plasma in W49B whose direct detection
has been recently reported. The initial conditions of the model were
set up according to the observational constraints derived for W49B and
for its inhomogeneous circumstellar environment.

Our model recovered the X-ray morphology of the remnant (the jet-like
feature and a sharp shell behind the molecular cloud highlighted by
XMM and Chandra observations; Miceli et al.\ 2006; Keohane et al.\
2007) and the spatial distribution of the overionized ejecta (localized
in the center of W49B and in its western region;
Miceli et al.\ 2010). We found that the thermal conduction plays an
important role in the evolution of the system, since it induces the
evaporation of a non-negligible fraction of the cold and dense cloud
material that mixes with the hot surrounding shocked medium, and promotes the
formation of the characteristic central bright feature of this mixed
morphology remnant (see Fig.~\ref{f:dentemp}).
Further more, as in W49B, similar coaxial rings and jet-like features are
also present in the X-ray emission in SNR 3C397 (rings in Jiang 2009;
jets in Safi-Harb et al.\ 2005 and Jiang \& Chen 2010) that is interacting
with a complicated molecular environment (Jiang et al.\ 2010); our
simulation may also provide a similar (but qualitative) explanation for the
structures in 3C397.

Our model predicts that overionization originates in the remnant due
to the rapid cooling of the hot plasma originally heated by the shock
reflected from the dense ring toward the center of the remnant. We
found two different classes of overionized plasma: one consists of
shocked interstellar medium (ISM), the other of a mixture of shocked ISM
and ejecta material. These two classes are associated with different
cooling processes, respectively: i) 
the mixing of hot shocked plasma with the cooler dense material evaporated from the ring.  
and ii) 
the rapid adiabatic expansion of plasma. In particular, the overionization
of the ejecta (observed in X-rays) is due to their rapid free expansion
that follows their early heating.

In this paper we adopted the simplest configuration of the initial
SN explosion, by considering a spherical symmetry. Nevertheless, this
simple configuration together with the inhomogeneities of the interstellar
medium reproduce the morphology and overionization pattern of the remnant
quite naturally. Our model suggests that the bipolar morphology of W49B
(e.g. its jet-like feature) could be an hydrodynamic effect originating
from the interaction of shocks reflected by the surrounding clouds. The
model indicates that the plasma in the remnant is in NEI, and it could
explain the unexpected Fe abundances found by Lopez et al. (2009) in
terms of overionization of the ejecta. It is worth emphasizing that
our analysis does not rule out the possibility of a bipolar explosion
in origin.  This alternative scenario deserve a further investigation
in a forthcoming paper.

On the basis of the morphology and conditions of overionization,
we derived $t\sim 2300$ yr as a good estimate for the age of W49B. Our
model shows that at this evolutionary stage, there is more overionized
plasma in the region where the ejecta expansion is not hampered. The
strength of the RRC of [Fe {\sc xxvi}] and its spatial distribution are
consistent with the X-ray observations. Our findings support, therefore,
the new idea that surrounding molecular clouds could be 
necessary ingredients
to reproduce the peculiar morphology, ionization conditions, and abundance
patterns of MM SNRs. High resolutions maps of RRC emission are needed to
confirm this scenario or other possible interpretations like in Yamaguchi
et al. (2009).

\section*{Acknowledgments}
We thank the anonymous referee for the valuable comments on our previous draft.
XZ would like to thank the friendly atmosphere at Osservatorio Astronomico
di Palermo. The software used in this work was in part developed by
the DOE-supported ASC / Alliance Center for Astrophysical Thermonuclear
Flashes at the University of Chicago, using modules for non-equilibrium
ionization, thermal conduction, and optically thin radiation built at
the Osservatorio Astronomico di Palermo. The simulation was performed
on the SCAN (Sistema di Calcolo per l'Astrofisica Numerica) HPC facility
of the INAF-Osservatorio Astronomico di Palermo. 
This work was partially supported by NSFC grants 10725312 and 10673003 and the China 973 Program grant 2009CB824800.
This work was partially supported by the ASI-INAF contract n.\ I/009/10/0.

\label{lastpage}

\end{document}